\documentclass{ws-ijmpe}
\usepackage[latin1]{inputenc}
\begin{document}

\markboth{R. Rodriguez-Guzman, J.L. Egido and L.M. Robledo}{
The superdeformed band of $^{36}$Ar}

%
\catchline{}{}{}{}{}
%

\title{Description of the superdeformed band of $^{36}$Ar with the Gogny force}

\author{R.R. Rodríguez-Guzmán}

\address{Institut f\"ur Theoretische Physik der
Universit\"at T\"ubingen,\\
Auf der Morgenstelle 14, D-72076 T\"ubingen, Germany.
}

\author{J.L. Egido and L.M. Robledo}

\address{Departamento de Física Teórica C-XI, Universidad Autónoma de Madrid\\
Madrid 28049, Spain\\
J.Luis.Egido@uam.es, Luis.Robledo@uam.es}

\maketitle

\begin{history}
\received{(received date)}
\revised{(revised date)}
\end{history}

\begin{abstract} 

The superdeformed band of $^{36}$Ar is studied with the Gogny force D1S and
the angular momentum projected generator coordinate method for the
quadrupole moment. The band head excitation energy, moments of inertia, 
$B(E2)$ transition probabilities and
stability against quadrupole fluctuations at low spin  are studied.  
The Self Consistent Cranking method is also used to describe the
superdeformed rotational band. In addition, properties of some 
normal deformed states are discussed.    
\end{abstract}


\section{Introduction.}

The lightest atomic nucleus known up to now to have  a superdeformed band (SD)
is  $ ^{36}$Ar \cite{Svensson.00}. The use of the sophisticated GAMMASPHERE
array detector has allowed the determination of the excitation energy of the
members of its SD band\cite{Svensson.00} up to $I=16\hbar$ (which is thought to
be the highest possible value of the band) as well as the in-band and some
out-band $B(E2)$ transition probabilities \cite{Svensson.01}. Theoretical
analyses indicate that the appearance of the SD band is linked to the
excitation of four particles from the $sd$ shell into the $fp$ one 
\cite{Svensson.00,Svensson.01,Ropke.02}. In addition, a normal deformed oblate
band is also known \cite{Endt.91}. In the theoretical side, this SD band has
been studied with the Cranked Nilsson-Strutinsky (CNS) and shell model
approaches \cite{Svensson.00,Svensson.01}, the Projected Shell Model (PSM)
\cite{Long.01} and with angular momentum projected techniques with the Skyrme
SLy6 interaction \cite{Bender.03}. 

The purpose of this paper is to study, using the Gogny interaction
\cite{Decharge.80} with the D1S parameterization \cite{Berger.84}, the
properties of the SD of the nucleus $ ^{36}$Ar  focusing on the stability of
the SD minimum against quadrupole fluctuations at low spin. The reason is that
in some of the theoretical studies mentioned  in the above paragraph the SD
minimum observed in the energy  landscape was very shallow rising serious
doubts about its ability to hold states at low angular momentum when
fluctuations in the quadrupole degree of freedom are taken into consideration
(see \cite{Moliqe.99} for a discussion of this issue). Obviously, at higher
spins the rotational energy makes the SD minimum deeper and therefore much more
stable against quadrupole fluctuations. As a side product of our calculations 
we have studied the properties of low lying normal deformed states and compared
them with the available experimental data. To perform the theoretical analysis,
we have used the Angular Momentum Projected Generator Coordinate Method
(AMP-GCM) with the axial quadrupole moment as generating coordinate and
restricted ourselves to $ K=0 $ configurations (see \cite{RRRG.02} for a
thorough discussion of the method and \cite{RRRG.00} for an application to the
study of SD bands). The properties of the superdeformed band obtained with the
AMP-GCM are also compared with those of a Self Consistent Cranking
calculation. 


\section{Theoretical framework.}

As we have already seen in many examples\cite{RRRG.02,RRRG.00}, whenever the
quadrupole degree of freedom plays a role in the nuclear dynamics it is
convenient to restore the broken rotational symmetry by projecting onto good
angular momentum. The effect of angular momentum projection in the energy
landscape can substantially change the mean field outcome and it turns out that
in many cases several minima appear with comparable energies. As a consequence,
configuration mixing of the quadrupole degree of freedom could be important. 
In this paper we have taken into account both angular momentum projection and
configuration mixing in the framework of the Angular Momentum Projected
Generator Coordinate Method (AMP-GCM). In this approach the wave functions of
the system are postulated to be 
\begin{equation} \label{eq1} 
\left| \Phi^{I}_{\sigma }\right\rangle = 
\int dq_{20}f^{I}_{\sigma}(q_{20})\hat{P}^{I}_{00}\left| \varphi (q_{20})\right\rangle. 
\end{equation}
In this expression $ \left| \varphi (q_{20})\right\rangle  $ is the set of
axially symmetric (i.e. $ K=0 $) Hartree-Fock-Bogoliubov (HFB) wave functions
generated with the constraint $ \left\langle \varphi (q_{20})\right|
z^{2}-1/2(x^{2}+y^{2})\left| \varphi (q_{20})\right\rangle =q_{20} $ on the
mass quadrupole moment. The HFB wave functions are expanded in an axially
symmetric Harmonic Oscillator (HO) basis with 10 major shells (220 HO states)
and equal oscillator lengths in order to preserve the rotational invariance of
the basis. In the evaluation of the HFB wave functions the Gogny force is used
(D1S parameterization). Other details concerning the HFB calculation are the
following: a) The two body kinetic energy correction is fully taken into
account in the variational process. b) The Coulomb exchange part of the
interaction, evaluated in the Slater approximation, has  been considered in the
variational process c) Reflection symmetry is imposed as a selfconsistent
symmetry of our HFB wave functions. 

The angular momentum projector operator 
\begin{equation}
\label{eq3}
\hat{P}^{I}_{00}=\frac{2I+1}{8\pi ^{2}}
\int d\Omega d^{I}_{00}(\beta )
e^{-i\alpha J_{z}}e^{-i\beta J_{y}}e^{-i\gamma J_{z}}
\end{equation}
is the usual one restricted to axially symmetric $K=0$ intrinsic 
configurations \cite{Ring.80,Hara.95} and $ f^{I}_{\sigma }(q_{20}) $
are the {}``collective amplitudes{}'' which are obtained as the solutions
of the Hill-Wheeler (HW) equation

\begin{equation}
\label{eq2}
\int dq'_{20}{\mathcal H}^{I}(q_{20},q'_{20})
f^{I}_{\sigma }(q'_{20})=
E^{I}_{\sigma }\int dq'_{20}{\mathcal N}^{I}(q_{20},q'_{20})
f^{I}_{\sigma }(q'_{20}).
\end{equation}

The solution of the HW equation for each value of the angular momentum $ I $
determines not only the ground state ($ \sigma =1) $, which is a member of the
Yrast band, but also excited states ($ \sigma =2,3,\ldots  $) that, in the
present context, may correspond to states with different deformation than the
ground state and/or quadrupole vibrational excitations. 

Further technical details pertaining the evaluation of the hamiltonian kernels
for density dependent forces and the evaluation of  transition probabilities
in the present framework are given in \cite{RRRG.02}.


\section{Discussion of the results.}


In Figure 1 we present the results of the HFB calculations used to generate the
intrinsic states $ \left| \varphi (q_{20})\right\rangle  $. On the left hand
side of the figure we show the HFB energy (panel (a) ) along with the $ q_{40}
$ deformation parameter (panel (b) ) and the particle-particle pairing  energy
$E_{pp}=-1/2Tr(\Delta \kappa ^{*}) $ (panel (c) ) for protons and  neutrons
(the two curves are indistinguishable) as a function of the quadrupole  moment
$ q_{20} $. The energy curve shows a deformed ground state minimum at $
q_{20}=-0.45 $b $ (\beta _{2}=-0.18) $ which is only  518 keV deeper than the
spherical configuration. A very shallow super deformed (SD) minimum at $
q_{20}=1.4 $b $ (\beta _{2}=0.52) $ is also observed at an excitation energy $
E^{HFB}_{x}(SD)=8.09 $MeV. The depth of the SD well is very small (only 67 keV)
and makes mandatory a study of the stability of such intrinsic state against
quadrupole fluctuations. To study the effect of the finite size of the basis in
the HFB energies and in the excitation energy of the SD minimum we have carried
out calculations including 18 major shells for the HO basis\footnote{This
basis is considered by some authors \cite{Moliqe.99} as almost
indistinguishable from an infinite basis for  nuclei around A=40.} (1140
states) for both the normal deformed (ND) and SD HFB minima and found that the
corresponding energies are shifted downwards by $ 906 $ keV and $ 1296 $ keV
respectively. As a consequence, the excitation energy of the SD minimum gets
reduced by $ 390 $ keV  up to the value $ E^{HFB}_{x}(SD)=7.7 $ MeV. 

\begin{figure*}
\includegraphics[width=0.99\textwidth]{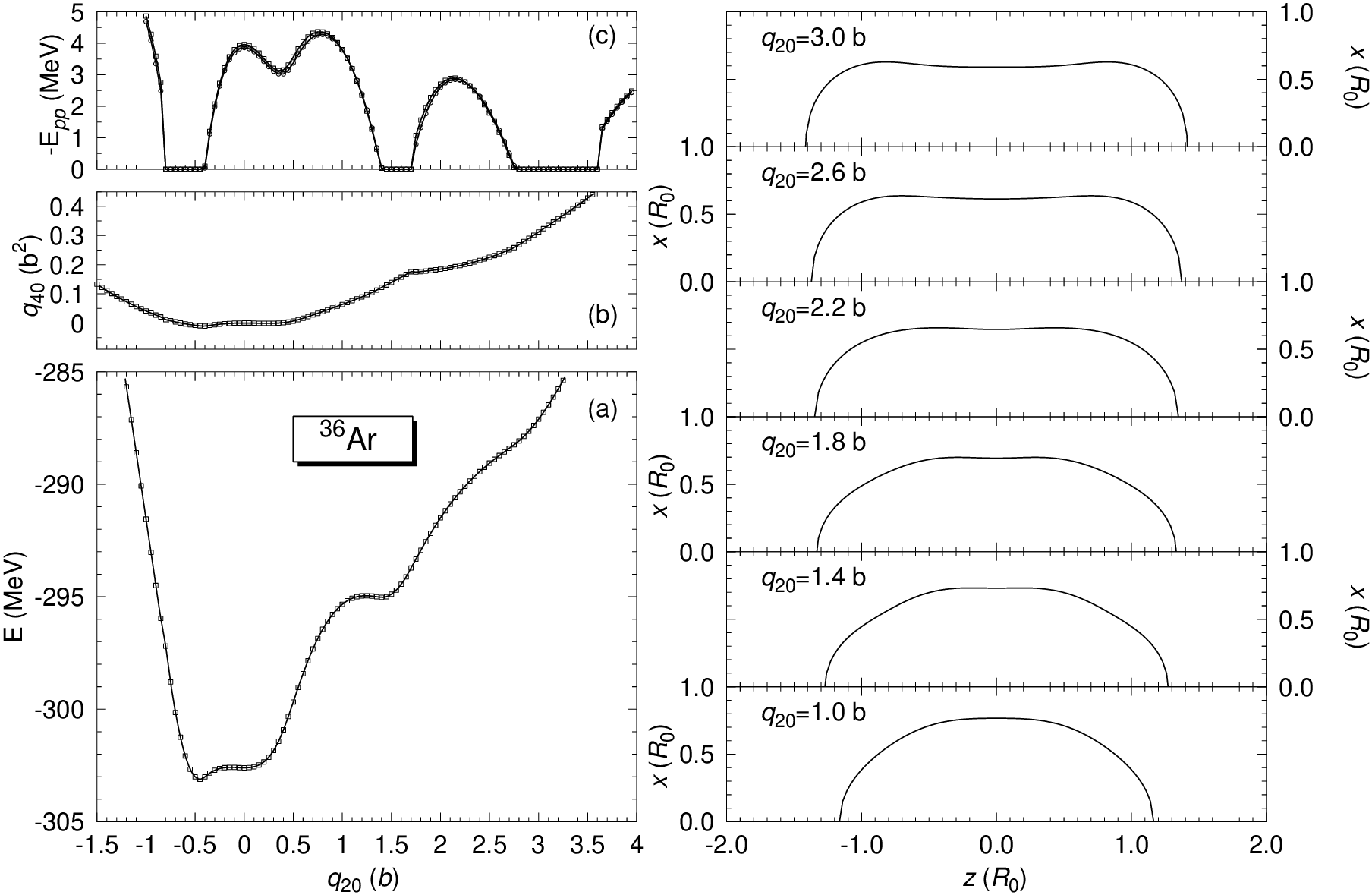}
\caption{
On the left hand side the HFB energy in MeV (panel (a) ), the \protect$ q
_{40}\protect $ deformation parameter in b$^2$ (panel (b) ) and the particle-particle
correlation energies \protect$ E_{pp}\protect $ for protons and neutrons in
MeV (panel
(c) ) are depicted as a function of the quadrupole moment \protect$
q_{20}\protect $ given in barns. On the right hand side, contour plots of the
matter distribution corresponding to a density of \protect$ 0.08 \textrm{fm}^{-3}\protect
$ and different quadrupole moments are depicted in units of
\protect$R_0=1.2A^{1/3} (3.96)\protect$ fm.}
\end{figure*}

The hexadecapole deformation parameter $ q_{40} $ increases with
increasing quadrupole moments and reach at the SD minimum the rather high value
$ \beta _{4}=0.24 $. Concerning the particle-particle correlation energies $
E_{pp} $ we observe that their values for protons and neutrons are nearly
identical and they go to zero in both the normal deformed and superdeformed
minima. This implies that dynamical pairing effects could be relevant for the
description of both the ND and SD bands. On the right hand side of the figure
we have plotted the matter density contour plots (at a density $ \rho
_{0}=0.08 \textrm{fm}^{-3} $) for several values of $ q_{20} $. Only for $
q_{20} $ values greater or equal 2.6 b the matter density distribution
resembles the one corresponding to two separated nuclei connected by  a rather
thick neck. Interestingly, this value of $ q_{20} $ corresponds to a shoulder
in the HFB energy landscape that will be analyzed later on when the effect of
angular momentum projection is considered.

To further study the physical contents of the ND and SD intrinsic wave 
functions we have computed the spherical shell occupancies $ \nu (nlj)=\sum
_{m}\left\langle \varphi \right| c^{+}_{nljm}c_{nljm}\left| \varphi
\right\rangle  $ for each  of these intrinsic  wave function. These quantities 
give the occupancy (or contents) of the HO orbital $ nlj $ in the intrinsic
wave function $ \left| \varphi \right\rangle  $. For the SD intrinsic state,
the positive parity  level occupancies are 5.26, 1.50 and 0.76 for the $
1d_{5/2} $, $ 1d_{3/2} $ and $ 2s_{1/2} $ orbitals, respectively, whereas for the
negative parity levels the occupancies are 1.05, 0.15, 0.83 and 0.18 for the $
1f_{7/2} $, $ 1f_{5/2} $, $ 2p_{3/2} $ and $ 2p_{1/2} $ orbitals, respectively
(the quantities for proton and neutrons are very similar and only the
neutron values are given). Therefore we have for the SD intrinsic state 15.04
particles in the $ N=2 $ major shell and 4.42 in the $ N=3 $ one in good
agreement with the shell-model results\cite{Svensson.00}.  It has to be
mentioned, however, that the $ 1d_{5/2} $ is not fully occupied in our
calculations in opposition to the shell model assumptions\cite{Svensson.00}. On
the other hand, the occupancy of the $fp$ shell in the ND minimum is negligible
as expected.

\begin{figure*}
\includegraphics[width=0.99\textwidth]{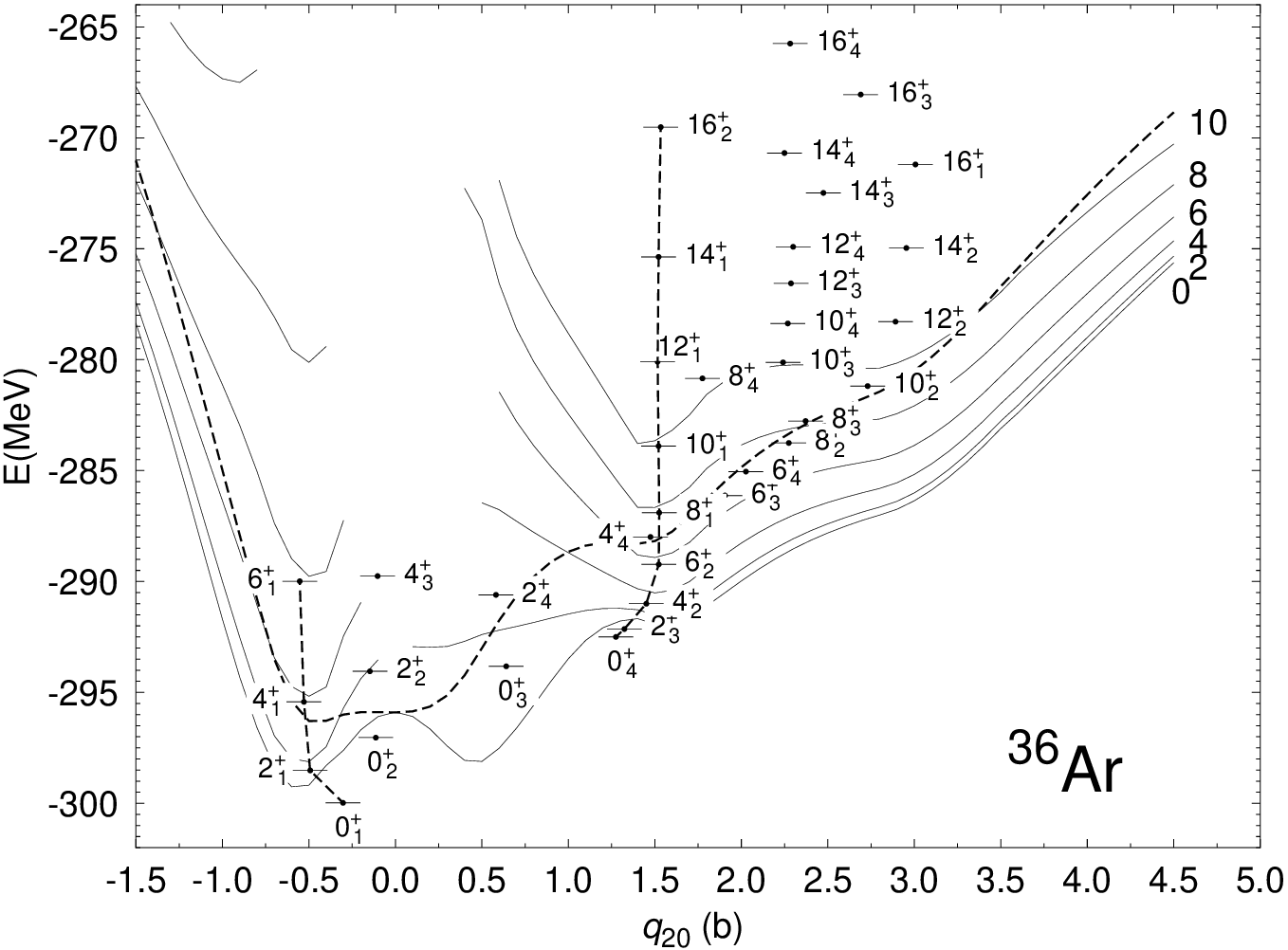}
\caption{The HFB energy (dashed line) and the angular momentum
projected energies up to \protect$ I=10\hbar \protect $ are plotted as a
function of the quadrupole deformation \protect$ q_{20}\protect $ measured
in barns. 
The four lowest-lying solutions of the AMP-GCM equation
are also plotted for each spin. The Coulomb exchange energy in the Slater
approximation has not
been added to the projected energy. See text for further details.}
\end{figure*}

In Figure 2 we have plotted the AMP energy curves for $ I=0,\ldots ,10\hbar  $
(full lines) along with the HFB energy curve (dashed line) as a function of the
quadrupole moment. The AMP $ I=0\hbar  $ energy curve shows more pronounced ND
and SD minima than the HFB one.  They are located at the quadrupole moments $
q_{20}=-0.55$ b and $ q_{20}=1.55$ b, respectively. The
excitation energy of the SD minimum with respect to the ground state for $
I=0\hbar  $ is $ E^{AMP}_{x}(SD)=7.37 $ MeV to be compared with the 8.09 MeV
obtained in the HFB calculation. Let us mention that performing the AMP
calculations with $ 18 $ shells is extremely time consuming and therefore we
will just use in this case the 390 keV shift obtained in the HFB to account for
the effect of the finite size of the basis in the excitation energy of the SD
band head. If we take into
consideration the 390 keV shift the excitation energy of the SD minimum in the
AMP case becomes $ E^{AMP}_{x}(SD)=6.98$MeV to be compared with the 7.7 MeV
obtained in the HFB case with 18 shells. We notice that for increasing spins,
the superdeformed minimum gets more and more pronounced and becomes the ground
state at spin $ I=8\hbar$.

We also show in Figure 2  the energies obtained in the AMP-GCM calculation for
the four lowest-lying solutions of the HW equation (labeled with the subindex $
\sigma =1,\ldots ,4 $) and spins from zero up to $ 16\hbar$. Each level has
been placed at a $ q_{20} $ value corresponding to its average intrinsic
deformation $(\overline{q}_{20})_\sigma^I$  (see \cite{RRRG.02} for its
definition). An oblate ND band
(states $0_1^+$, $2_1^+$, $4_1^+$ and $6_1^+$, joined by a dotted line) is
clearly observed together with a SD band ($0_4^+$, $2_3^+$, $4_2^+$, $6_2^+$,
$8_1^+$, etc) that becomes Yrast at angular momentum 8 $\hbar$. 
Many hyperdeformed states are also observed for spins greater of equal 8 and
their appearance is linked to the shoulder observed in the HFB energy
landscape at $q_{20}=2.6$ b that becomes a minimum in the AMP energy landscapes
at $I=8\hbar$.

The binding energy obtained for the AMP-GCM (adding the Coulomb exchange energy
in the Slater approximation and taking into account the 906 keV shift due to
the finite size of the basis) is 307.606 MeV in very good agreement with the
experimental value of 306.715 MeV. This result has to be compared with the
HFB value of 304.020 MeV.

The ND band results compare well with the experimental data (see Table 1) both for the
excitation energies and $B(E2)$ transition probabilities.

\begin{table*}
\begin{center}
\begin{tabular}{c|c|c||c|c|c}\hline 
\multicolumn{3}{c||}{Experiment}& \multicolumn{3}{c}{Theory} \\ \hline  \hline 
$ J^{\pi } $ &  E (MeV)  & $ B(E2)\downarrow  (e^{2}\textrm{fm}^{4})$ & 
$ J^{\pi } $ &  E  (MeV) & $ B(E2)\downarrow  (e^{2}\textrm{fm}^{4})$ \\ \hline \hline
$ 0^{+} $    & 0.00      & --         & $ 0_{1}^{+} $ &  0.00 & --  \\ \hline
$ 2^{+} $    & 1.97      & 60 $\pm$ 6 & $ 2_{1}^{+} $ &  1.45 & 72.1  \\ \hline
$ 4^{+} $    & 4.41      &            & $ 4_{1}^{+} $ &  4.54 & 102.3  \\ \hline
$ 6^{+} $    & 9.18      &            & $ 6_{1}^{+} $ & 10.01 & 112.8 \\ \hline
\end{tabular}
\end{center}
\caption{
Excitation energies and \protect$ B(E2,I\rightarrow I-2)\protect $
values of the normal deformed oblate band.}
\end{table*}

Concerning the SD band the first noticeable fact is that, in spite of the small
depth of the SD well, the quadrupole fluctuations preserve the SD intrinsic
state. Obviously the mixing with other quadrupole configurations reduces the
deformation of the $0^+_4$ intrinsic state as compared to the one of the HFB
method but not substantially. As a consequence of configuration mixing the
excitation energy of the SD band becomes 7.476 MeV (7.086 MeV including the 390
keV shift due to the finite size of the basis) which is far from the
experimental value of 4.33 MeV. It is not clear whether  the discrepancy should
be attributed to the interaction used (rather unlikely, as the same kind of
disagreement is also obtained with the Skyrme SLy6  interaction
\cite{Bender.03} where the SD band-head is predicted at 5.9 MeV)  or to other
degrees of freedom not taken into  account like triaxiality or  even
octupolarity.

Coming back to the SD band, the calculated excitation energies compare pretty 
well with experiment up to the back-bending observed at $I=10\hbar$ but the
$B(E2)\downarrow$ transition probabilities come up too high. The result for the
$B(E2)$ can be understood by looking at the results of the CNS calculations of
ref. \cite{Svensson.00} where it is observed that already at moderate spins the
SD band members become triaxial and their deformation decreases as a function
of spin. In our calculations, which do not include triaxiality, the SD band
intrinsic  deformation remains constant from spin 6 $\hbar$ on explaining the
rigid rotor behavior for the $B(E2)$'s. To study the effect of triaxiality we
have performed Self Consistent Cranking (SCC) calculations for the SD band and
observed rather strong triaxiality effects even at zero spin and a steady
decrease of deformation with increasing angular momentum (from $\beta_2=0.52$
at zero spin down to $\beta_2=0.4$ at spin 16 $\hbar$). 

\begin{figure*}
\includegraphics[width=0.99\textwidth]{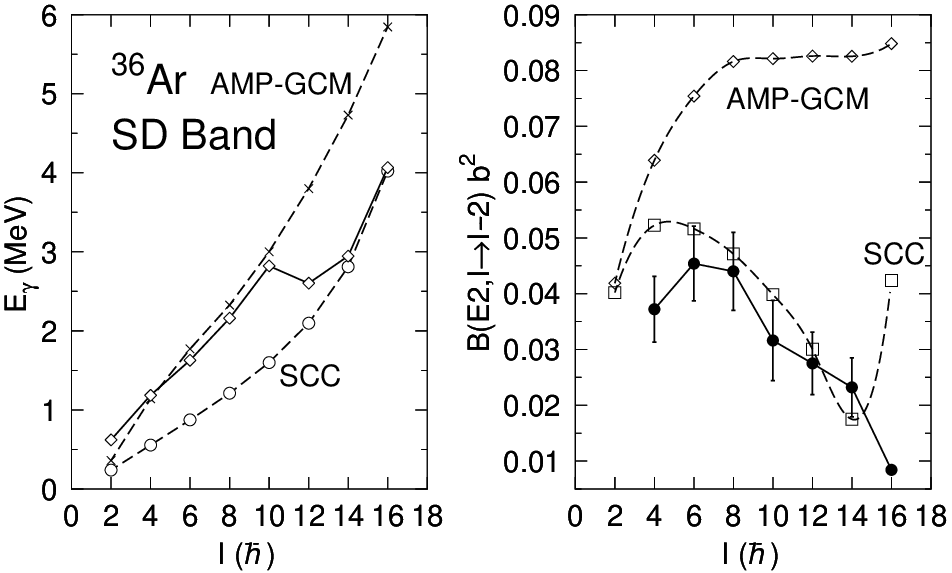}%
\caption{In the left hand side the gamma ray energies \protect$ E_{\gamma }(I)=E(I)-E(I-2)\protect $
in  MeV for the SD configuration are depicted as a function
of spin \protect$ I(\hbar )\protect $ for the  AMP-GCM and SCC calculations
(dashed lines) along with the experimental results (full line).
In the right hand side  both the theoretical (AMP-GCM and SCC) and experimental results for
the \protect$ B(E2,I\rightarrow I-2)\protect $ transition probabilities in
units of \protect$ e^{2}\textrm{b}^{2}\protect $ are depicted as a function of spin.}
\end{figure*}

In Fig. 3 we have
summarized the results of both the AMP-GCM and SCC calculations and compared
them with the experimental data. We observe that the SCC results for the last two members of the
SD band agree well with the experiment but for lower spins the theoretical
$\gamma$ ray energies are too low. This is a consequence of the lack of dynamical
pairing in our calculations. Pairing correlations are zero at the SD intrinsic
configuration in our calculations producing a too high moment of inertia (see
\cite{Long.01} for a discussion of the pairing properties of this SD band). It is
to be expected that the inclusion of dynamical pairing in our calculations 
will reduce the moment of inertia substantially bringing theory and experiment
in much better agreement. At high spins it is expected that pairing
correlations will be strongly suppressed by the Coriolis anti-pairing effect and
therefore our no-pairing results agree well with the experiment in that region.
Concerning the $B(E2)$ transition probabilities we observe a nice agreement
between the SCC results and the experiment as well as the strong discrepancy
between the AMP-GCM values and the experimental results.

From the comparison of the $B(E2)$ values it is clear that the SCC is far
superior than the AMP-GCM method for the description of the SD band as it takes
into account triaxial effects responsible for the decreasing of deformation
with increasing spins. Therefore, we can consider that the nice agreement
between the AMP-GCM excitation energies and the experiment is accidental and
probably due to the fact that further improvements of the AMP-GCM will  cancel
out: in one hand, the dynamical pairing correlations that should decrease the
moment of inertia and in the other the consideration of $\Delta K \ne 0$
admixtures in the intrinsic wave function that should increase the moment of
inertia (see the discussion of this effect in \cite{RRRG.00b}). However, in spite of
the above deficiencies, it has to be pointed out that one of the merits of the
AMP-GCM is that is the  only method able to asses the stability of the SD
intrinsic configuration at low spins.

\section{Conclusions}

We have analyzed the ND and SD bands of $^{36}$Ar in the framework of the Angular
Momentum Projected Generator Coordinate Method (AMP-GCM) and the Self Consistent
Cranking (SCC) approach. The AMP-GCM permit us to asses the stability against
quadrupole deformation of the SD minimum at low spin and also allows a nice
description of the ND band. For the SD band, the expected triaxiality downgrades
the quality of the AMP-GCM description but we have found that the SCC is in this
case rather satisfactory.

\section{Acknowledgments.}

This work has been supported in part by the DGI, Ministerio de Ciencia y
Tecnología (Spain), under project  BFM2001-0184.

\end{document}